\documentclass{amsart}
\usepackage{amssymb,latexsym}
\def\bo{{
\hfill $\Box$ }}

\newcommand{\ep }{\varepsilon}
\newcommand{\vp }{\varphi}
\newcommand{\bm}{\boldmath}
\newcommand{\ubm}{\unboldmath}
\newcommand{\bR}{\mbox{\bm $R$\ubm}}
\newcommand{\bRn}{\bR^n}
\newcommand{\bC}{\mbox{\bm $C$\ubm}}
\newcommand{\bN}{\mbox{\bm $N$\ubm}}
\newcommand{\bCn}{\bC^n}
\newcommand{\pard}{\partial}
\newcommand{\pardi}{\pard_i}
\newcommand{\pardj}{\pard_j}
\newcommand{\pardN}{\pard_N}
\newcommand{\leftab}{\left\arrowvert}
\newcommand{\rghtab}{\right\arrowvert}

\newcommand{\eq}[1]{Eq.~(\ref{#1})}

\newcommand{\num}[1]{(\ref{#1})}
\newcommand{\Hloc}[1]{H^{#1}_{\rm{loc}}(D')}
\newcommand{\Lloc}{L^{2}_{\rm{loc}}(D')}

\newtheorem{defn}{Definition}[section]
\newtheorem{theorem}{Theorem}[section]
\newtheorem{prop}{Proposition}[section]
\newtheorem{lemma}{Lemma}[section]

\begin{document}
\title{Existence and uniqueness of the scattering solutions \\
in the exterior of rough domains }

\author{ Alexander G. Ramm}
\address{ Department of Mathematics\\
Kansas State University\\
Manhattan, KS 66506-2602, USA.\\
Email: ramm@math.ksu.edu}
\author{Marco Sammartino}
\address{
Dipartimento di Matematica,\\
Via Archirafi 34, 90123 \\
Palermo, Italy.          \\
E-mail: marco@dipmat.math.unipa.it
}

\begin{abstract}
A simple and short proof is given for the existence
and uniqueness of the solution to the obstacle scattering problem
under weak smoothness assumptions on the obstacle.
The Robin, Neumann and Dirichlet boundary conditions are considered.
\end{abstract}


\maketitle

\vspace{.5cm}
\section{Introduction}
\setcounter{equation}{0}

In this paper we study the scattering problem in the
exterior of a rough  bounded domain.
This  problem was investigated in \cite{RR}, where
it was assumed that the potential had  compact support.
The goal of this paper is to relax this hypothesis
and to give a simple and general method of proof which
is simpler than the earlier known. The assumptions on the coefficients
of the differential operator are also relaxed.
We  prove the results, similar to those in \cite{RR},
assuming
that the potential decays at infinity at the power
rate which  depends on the space dimension.
We discuss the 3-D case ($n=3$),
but the arguments are similar in the $n$-dimensional case.
By this reason in many places we keep $n$ in the formulation
of the results and assumptions. Estimates (3.18), (3.19), (4.1), (4.2)
and (4.16) are given for $n=3$.

The assumptions on the smoothness of the
domain are minimal and include all the previously studied
cases and probably all of the cases of interest in applications.

The plan of the paper is as follows. In the rest of this Section
we introduce some notations, the function spaces we work with,
and give  the statement of the scattering problem.
In Section 2 we prove the uniqueness of the solution of the
scattering problem.
In Section 3 we prove the existence of the solution
assuming that the potential has  compact support.
In Section 4 we
assume a power decay rate of the potential and prove the existence
of the scattering solution.

Our results are stated and proved for the broader class of
domains than in the literature, see e.g. \cite{MMP}, \cite{M} and
\cite{V}. In particular Lipschitz domains form a proper subset in the
class of
domains we study. An inverse obstacle scattering
problem in a class of rough domains
was studied in \cite{R3}.

The basic ideas of our method are simple: first, we prove
that the operator of the problem under consideration is
selfadjoint. This is done by using the fact that every closed symmetric
densely defined semibounded from below quadratic form on a Hilbert
space defines a unique selfadjoint operator whose domain is dense in the
domain of the quadratic form.
The assumptions (1.4) and (1.5) (see below) concerning
the roughness of the domain guarantee that the corresponding quadratic form
is semibounded from below and closed.
The case of the Dirichlet boundary condition is the simplest one,
and in this paper we mostly deal with the Robin boundary condition.
In the simplest case of
the Dirichlet boundary condition no assumptions are needed on the
roughness of the obstacle: any compact domain is admissible. The reason
is simple: for the compactness of the embedding $H^1_0(D_R)\to L^2(D_R)$
no assumptions on $D$, except boundedness of $D$ are needed, while
for the compactness of the embedding $H^1(D_R)\to L^2(D_R)$
(see next section for the notations and condition (1.4) below)
some assumptions concerning the roughness of the boundary
of $D$ are necessary. Necessary and sufficient conditions
for the compactness of the above embedding operator are
known \cite {Ma}, \cite {MaP}. For the Robin boundary condition
in addition to (1.4) assumption (1.5) is needed.

Secondly, we prove that the solution
to the problem with complex spectral parameter $k^2+i\epsilon$,
$\epsilon >0$, converges, as $\epsilon \to 0$, to the solution
of the scattering problem. The convergence holds in suitable local
and global norms.

The underlying idea is to use the Fredholm
property of the problem in a properly chosen pair of normed spaces.

\subsection{Notations and assumptions}

The following are the notations used in this paper.

\smallskip\noindent
By $D\subset \bRn $ we denote a bounded domain,
      $D'$ stands for its complement, $D':=\bRn\setminus D$,
      $D_R:=D'\cap B_R$,
      $B_R$ denotes a ball with radius $R$
      such that $B_R\supset D$, and
      $B'_R$ is  the complement of $B_R$ in $\bRn$.

\smallskip\noindent
By $a_{ij}(x)$, $x\in D'$, we denote the elements of
      a real-valued symmetric matrix satisfying the
      following hypotheses:
\begin{equation} \exists c,\, C >0 \;\; \mbox{such that }
\qquad c|t|^2\leq a_{ij}(x)t_i \bar{t}_j\leq C|t|^2 \qquad \forall t\in
\bCn\; , \forall x\in D'\; .  \label{1.1}
\end{equation}
In (\ref{1.1}) and below
the summation over the repeated indices is understood. One has:
\begin{equation}
a_{ij}(x)=\delta_{ij} \;\;\mbox{when} \;\; |x|>R \; , \label{1.2}
\end{equation}
and the coefficients $a_{ij}(x)$ are assumed Lipschitz continuous. This
assumption implies the validity of the unique continuation principle for
the solutions to elliptic equation (\ref{1.sc}) below.

\smallskip\noindent
Let
$$
lu:=-\pardj\left( a_{ij}(x)\pardi u\right) \; .
$$

\smallskip\noindent
Assume
\begin{equation}
\begin{array}{l}
q(x)=\overline{q(x)},\quad q(x)\in L_{\rm{loc}}^p (\bRn),\quad p >
\displaystyle\frac{n}{2},\quad q(x)\in L^{\infty}(\tilde S),
\\[1.5ex]  \displaystyle
|q(x)|\leq \frac{c}{(1+|x|^2)^{s/2}},\quad |x|>R,\ s>n, \;
n\geq 3,
\end{array}
\label{1.3}
\end{equation}
where $\tilde S \subset D'$ is a neighborhood of the boundary
$S:=\partial D$.
The bar stands for complex conjugate. The assumption $s>n$
is weakened in some of the statements of
this paper: for example, in
Proposition 2.2 below it is assumed that $s>1$.
However, the assumption $s>n,\, n=3$, is used in section 4.
In \cite{A} the scattering theory
is developed under the assumptions
that $a_{ij}=\delta_{ij}$, the obstacle
is absent, and $q(x)$ satisfies less restrictive assumptions than
(1.3), essentially it is assumed that $s>1$. Therefore it is
likely that the results of our paper can be obtained under this weaker
assumption. However, much additional technical work is needed
to obtain such results. We are emphasizing the methodology of
handling rough boundaries in this paper and discussing
of the wider class of potentials would lead us astray.

\smallskip\noindent
Let
$$
Lu:=lu+q(x) u \; .
$$

\smallskip\noindent
Let $S:=\partial D$. Our main assumptions
concerning the smoothness of $D$ are:
\begin{equation} \mbox{The imbedding} \quad
i:H^1(D_R)\rightarrow L^2(D_R)\quad\; \mbox{ is compact}\; , \label{1.4}
\end{equation} \
\begin{equation}
\mbox{The trace operator} \quad r:H^1(D_R)\rightarrow
L^2(S)\quad\; \mbox{ is compact}\; , \label{1.5}
\end{equation}
where in the definition of $L^2(S)$  the integration over $S$
is understood with respect to the
$n-1$-dimensional Hausdorff measure \cite {Ma}.
In particular, assumption (1.5) implies that
the $n-1$-dimensional Hausdorff measure of $S$ is finite,
that $D$ has a finite perimeter (see \cite {Ma}, p.296),
and that Gauss-Green formula holds in $D$ for
functions in $BV(D)$ space, which consists
of functions $u\in L^1_{loc}(D)$ such that $\nabla u$,
understood in the sense of distribution theory,
is a signed measure (a charge) in $D$.
See \cite{Ma} for the exact formulation of
the Gauss-Green formula for such domains and
such class of functions.

Let
$$
u_N:=\pardN u:=a_{ij}(x)\pardi u N_j \; ,
$$
where $N_j$ is the j-$\it{th}$ component of the normal
to $S$ pointing into $D$.
The definition of the normal for rough domains is not discussed
here because in our formulation of the scattering problem
(see (1.11)--(1.12) below) the notion of normal is not used.
One can find the definition of the normal
in the sense of Federer in \cite{Ma}, p.303.

In this paper we do not define the class of rough domains
explicitly, but isolate property (1.4) or
properties (1.4) and (1.5) as defining properties.
In \cite{Ma} necessary and sufficient conditions on $S$,
the boundary of $D$, are given for these properties to hold.
In formulas below in which the surface integrals
appear, e.g., (1.11), (2.1), (3.1), (3.2), etc,
the integration measure $ds$ is the $n-1$-dimensional Hausdorff measure
defined, for example in \cite {Ma} p.37.
In \cite {Fed} it is proved that for rectifiable surfaces
(the class of these surfaces is much broader than the class
of Lipschitz surfaces) the
$n-1$-dimensional Hausdorff measure is equivalent to the
measure generated by the elements of the surface area.

\smallskip\noindent
Given an $L^{\infty}(S)$ real-valued function $\sigma(s)$, $s\in S$,
one defines the operator $\Gamma^R$:
$$
\Gamma^R u(s):= u_N(s)+\sigma(s) u(s), \quad s\in S\; .
$$
      Introduce the following operators:
\begin{eqnarray}
\Gamma^D u(s) &:=&  u(s),  \quad s\in S\; ,   \nonumber \\
\Gamma^N u(s) &:=& u_N(s), \quad s\in S\; .   \nonumber
\end{eqnarray}

\smallskip\noindent
Let   $u_0:=\exp{(ik\alpha\cdot x)}$,  $\alpha\in S^{n-1}$, where
$S^{n-1}$ is the unit
sphere in $\bf {R^n}$, $k>0$, and
$$
v:=u-u_0 \; .
$$

\subsection{Function spaces}

Let
\begin{equation}
(u,v):=\int_{D'}u\bar{v}dx  \; . \label{}
\end{equation}
On $H^1\left(D'\right)$ define the following bilinear form,
$[\cdot, \cdot]: H^1\left(D'\right)\times H^1\left(D'\right)
\rightarrow \bC$:
\begin{equation}
[u,v]:= \int_{D'} a_{ij}\pardi u \pardj \bar{v} dx + (u,v)
\; . \label{biquad}
\end{equation}
We use  the following spaces:
\begin{defn}
The space $\Lloc$ is the space of  functions  $f$ such that,
$\forall r>R$,  $f\in L^2(D_r)$.
\end{defn}
\begin{defn}
The space $\Hloc{l}$ is the space of  functions  $f$ such that,
$\forall r>R$,  $f\in H^l(D_r)$, where $H^l(D_r)$
is the Sobolev space.
\end{defn}
\begin{defn}
The space $H^1_0(D')$ is the closure in $H^1(D')$ norm of the space
 of functions 
  $f\in H^1(D')$
vanishing near $S$. The set of functions which belong to
$H^1_0(D_r)$ for any $r>R$ is denoted by $H^1_{0 loc}(D')$.
\end{defn}
\begin{defn}
The space $\tilde{H}^1(D')$ is the space of  functions $f\in H^1(D')$
vanishing near infinity.
\end{defn}
\begin{defn}
The space $\tilde{H}^1_0(D')$ is the space of  functions
$f\in H_0^1(D')$
vanishing near infinity.
\end{defn}
\begin{defn}
The space $H^{\ell}_s:=H^{\ell}_s(D'),$
$\ell=0,1,$ is the space of functions
$f\in \Hloc{\ell}$ with finite
weighted $L^2(D')$-norm, for example the norm in $H^{\ell}_s$
with $\ell=1$ is defined as follows:
$$
\| f\|^2_{1,s}=\int_{D'}\left(1+|x|^2\right)^{s/2}\left( |f|^2+
|\nabla f|^2\right)dx <\infty \; .
$$
\end{defn}

\subsection{Statement of the problem }

We study the following
problem:
\begin{eqnarray}
&& Lu-k^2 u   
=0\quad\mbox{in}\;  D' \; ,   \label{1.sc}  \\
&& \Gamma^X u 
=0\quad\mbox{in}\;  S  \; ,    \label{1.bo }  \\
&& \lim_{r\rightarrow \infty}\int_{|x|=r} |v_r -ikv|^2 ds
=0, 
\label{1.ra}
\end{eqnarray}
where $v=u-u_0$, $\;u_0=e^{i k\alpha\cdot x}$,
$\alpha \in S^{n-1}$ is given, $k>0$ is fixed, and
$X=R$ (the Robin boundary condition), or
$X=N$ (the Neumann boundary condition), or
$X=D$ (the Dirichlet boundary condition).
We mostly discuss the Robin boundary condition.
The other cases can be treated similarly.
The Neumann boundary condition is a particular case of
the Robin boundary condition ($\sigma(s)\equiv 0$).
The Dirichlet boundary condition is the simplest: it does not require
any smoothness
assumptions concerning the boundary $S$,
only the boundedness of $D$ is required.
For the Neumann boundary condition only assumption
\num{1.4} is needed.
For the Robin boundary condition both assumptions \num{1.4}
 and \num{1.5} are used.

\subsection{The weak formulation of the scattering problem}

Here we introduce the weak formulation of  problem
\num{1.sc}--\num{1.ra}.

\begin{defn}
We say that $u\in H^1_{loc}( D')\cap H^1_{-s}(D'), \, s>1,$ is a
weak solution
of the scattering problem (\ref{1.sc})--\num{1.ra} with $X=R$
(the Robin boundary condition) if
\begin{eqnarray}
&&\int_{D'}\left[ a_{ij}\pardi u \pardj \bar{\vp} +
\left(q-k^2\right) u\bar{\vp} \right] dx +\int_S \sigma u\bar{\vp}ds
=0
 \quad \forall \vp\in \tilde{H}^1\left( D'\right),
 \label{1.8} \\
&&\lim_{r\rightarrow \infty}\int_{|x|=r} |v_r -ikv|^2ds
=0,
\;
\quad \mbox{\rm{where}}\quad v=u-u_0  \; .
\end{eqnarray}
\end{defn}

\begin{defn}
We say that $u\in H^1_{0\,loc}\left(
D'\right)\cap H^1_{-s}(D'), \, s>1,$ is a weak solution
of the scattering problem (\ref{1.sc})--\num{1.ra} with $X=D$
(the Dirichlet boundary condition) if
\begin{eqnarray}
&&
\int_{D'}\left[ a_{ij}\pardi u \pardj \bar{\vp} +
\left(q-k^2\right) u\bar{\vp} \right] dx
=0
\quad
\forall \vp\in \tilde{H}^1_0\left( D'\right)  \; ,
\\
&&\lim_{r\rightarrow \infty}\int_{|x|=r} |v_r -ikv|^2ds
=0,
\;
\quad \mbox{\rm{where}}\quad
v=u-u_0 \; .
\end{eqnarray}
\end{defn}

\section{Uniqueness theorem}
\setcounter{equation}{0}

In this section we  prove uniqueness of the solution to problem
(\ref{1.sc})--\num{1.ra}.
We first state the following known results:

\begin{prop}\label{p2.1} (\cite{W}, p.227)
Suppose $u$ is a solution of \eq{1.sc}
vanishing on an open subset of $D'$.
If $a_{ij}(x)$ are Lipschitz and $q(x)\in L^p_{loc}(R^n),
p>\frac n2, n\geq 3$,
then $u\equiv 0 $ in $D'$.

\end{prop}
The above Proposition is called the {\it
unique continuation principle.}
\begin{prop} \label{p2.2} (\cite {Ramm},  p.25)
Suppose that (\ref{1.2}) and (\ref{1.3}) hold,
$s>1$ in (\ref{1.3}) and $k>0$. If
$u\in H^1_{loc}(B_R')$ and
\begin{eqnarray}
&&\int_{ B'_R}\left[ \pardi u\pardi \bar{\vp} +
\left(q(x)-k^2\right)u\bar{\vp}\right]dx
=0
\quad\;
\forall \vp\in \tilde H^1_0(B'_R), \label{2.1} \\
&&\lim_{r\rightarrow\infty}\int_{|x|=r}|u|^2 ds
=0
, \label{2.2}
\end{eqnarray}
then $u\equiv 0$ outside $B_R$.
\end{prop}

In \cite {Ramm}, p.25, it is assumed that $q(x)=0$. In
the general case the result follows from a theorem of
T.Kato (\cite {K}) which says that any solution
to equation (1.8) in $B'_R$, which satisfies (2.2),
vanishes in $B'_R$ if $k>0$ and $|x||q(x)|\to 0$ as $|x|\to \infty$.
It is not necessary to assume $q$ real-valued in Kato's theorem.

We now prove a Lemma used  in the proof of the
uniqueness of the solution of the scattering problem.
\begin{lemma} \label{l2.1}
Suppose $W$ satisfies \num{1.8}.  Then
\begin{equation}
\int_{|x|=r} \left( \bar{W}W_r-W\bar{W}_r\right)ds =0\;, \quad r>R.
\end{equation}
\end{lemma}
{\bf Proof}: Take a cut-off
 function $h(r)\in C^\infty(\bR)$ such that $0\leq h \leq 1$,
 $h(r)=1$ when $r\leq 1/2$, $h(r)=0$ when $r\geq 3/2$,
$h$ is monotonically decreasing.
In \num{1.8} take $\bar{\vp}=\bar{W}h\left((|x|-r_0)/\delta\right)$,
with $r_0>R$, and $\delta>0$.
Then take the complex conjugate and subtract. Using (\ref{1.2}),  one gets:
$$
\int_{D'}\left( \bar{W}\pardi W  - W\pardi \bar{W}  \right)
\pardi h\left((|x|-r_0)/\delta\right)dx =0\; .
$$
If one takes the limit $\delta\rightarrow 0$ in the above relation,
one gets the desired result. Taking this limit,
one uses the interior regularity of $W$: this regularity
is a consequence of equation
\num{1.8} and of the
 known interior elliptic
regularity results.

We now state the main result of this Section.
\begin{theorem} \label{th2.1}
Suppose $u_1$ and $u_2$ are the
 weak solutions of the scattering problem
with the Robin (Neumann, Dirichlet) boundary condition. Then $u_1\equiv
u_2$.
\end{theorem}
{\bf Proof}: We prove this Theorem in the case of the
Robin boundary condition. The cases of the
Neumann and the Dirichlet boundary conditions can be treated similarly.
Define $W:=u_1-u_2$. One has:
\begin{eqnarray}
&&
\int_{D'}\left[ a_{ij}\pardi W \pardj \bar{\vp} +
\left(q-k^2\right) W\bar{\vp} \right] dx +\int_S \sigma W\bar{\vp}ds
=0
  \quad  \forall \vp\in \tilde{H}^1\left( D'\right),
\label{scatter}
\\
&&
\lim_{r\rightarrow \infty}\int_{|x|=r} |W_r -ikW|^2ds
=0
\; .
\label{radiat}
\end{eqnarray}
Equation \num{radiat} can be written as:
$$
\lim_{r\rightarrow \infty}\int_{|x|=r} \left(|W_r|^2+k^2|W|^2\right)ds+
\lim_{r\rightarrow \infty}ik\int_{|x|=r}
\left(W_r\bar{W}-\bar{W}_rW\right)ds =0\; .
$$
Since $W$ satisfies \num{scatter}, Lemma \ref{l2.1} applies, and
the above relation implies
$$
\lim_{r\rightarrow \infty}\int_{|x|=r} |W|^2 ds=0\; ,
$$
which is condition \num{2.2} of Proposition \ref{p2.2}.
Moreover, from \num{scatter} and \num{1.2} it follows
that $W$ satisfies condition \num{2.1}.
Therefore Proposition \ref{p2.2}
implies $W=0$
outside $B_a$.
Then one applies Proposition \ref{p2.1} and concludes that $W\equiv 0$
in $D'$.
Theorem~\ref{th2.1} is proved.
\bo

\vskip.3cm
We wish to prove the
existence of the solution of the scattering problem.
Let us reduce the problem to the one for a function
which satisfies the radiation condition at infinity.

Take a cut-off function $\zeta\in C^\infty(\bR)$,
such that $0\leq \zeta\leq 1$,
$\zeta\equiv 0$ in a neighborhood of $D$ and
$\zeta\equiv 1$ outside $B_r$ for some $r>R$, and  define
$$w:=u-\zeta u_0.$$
If $u$ solves the
scattering problem \num{1.sc}--\num{1.ra}, then $w$ solves
the following problem:
\begin{eqnarray}
&&
Lw-k^2 w   
=
f:=\left(L-k^2\right)(\zeta u_0), \label{2.6} \\[0.5ex]
&&\Gamma^X w 
=0, \label{2.7} \\[0.5ex]
&&\lim_{r\rightarrow \infty}\int_{|x|=r} |w_r -ikw|^2ds 
=0.
\label{2.8}
\end{eqnarray}
In Section 3 we prove that the above problem has  a
solution.
Theorem~\ref{th2.1} implies that this solution is unique.

\section{Existence for  compactly supported potential}
\setcounter{equation}{0}

In this Section we  prove that the scattering problem
\num{1.sc}--\num{1.ra} with a compactly supported potential,
 $q(x)=0$ if  $|x|>a$,
has the  unique solution.
We look for a solution of the form $u=w+\zeta u_0$, where
$\zeta$ is a cut-off function (defined above formula (2.6))
and $w$ satisfies \num{2.6}--\num{2.8}.
Let us prove that  problem \num{2.6}--\num{2.8}
has  a solution.

\subsection{Existence for the equation with the absorption}

Consider the weak form of the
scattering problem:

Let $\ep>0$.
Find $w\in H^1_{loc}\left( D'\right)\cap H^1_{-s}(D'),\, s>1,$ such that:
\begin{equation}
\qquad\int_{D'} \left[a_{ij}\pardi w\pardj \bar{\vp} dx +
\left(q-k^2-i\ep\right)w\bar{\vp} \right] +
\int_S \sigma w\bar{\vp}ds
=(f,\vp)\;
\quad \forall \vp\in \tilde{H}^1\left( D'\right),
\label{3.1}\;
\end{equation}

In the space $H^1(D')\subset L^2(D')$
consider the following bilinear  form:
\begin{equation}
B_\gamma[u,v]:=[u,v]+\gamma(u,v)+\int_S\sigma u\bar v dS
+(qu,v):=[u,v]_\gamma+(qu,v) 
\, ,
\end{equation}
where $[u,v]$ is defined in \num{biquad}
and $\gamma >0$ is chosen so large that the
form $[u,u]_\gamma^{1/2}$ defines the norm equivalent to $H^1(D')$.
We use here assumption (1.5), which implies that the
boundary integral in (3.2) can be estimated by the
term
\linebreak
$c ||u||_{H^1(D_R)}||v||_{H^1(D_R)}$.
One could omit assumption (1.5) and change the space
in which the solution is sought,  to the space
with the norm containing
the additional term $||u||_{L^2(S)}$, where
the measure used in the definition of $L^2(S)$ is the
$n-1$-dimensional Hausdorff measure. Then one could
use assumption (1.4), do not use
assumption (1.5), and assume that $S$ has finite
perimeter (see \cite {VH}). A set has finite perimeter
if the gradient (in the sense of distribution
theory) of the characteristic function of this set
belongs to the space $BV(D)$. This space was mentioned
below formula (1.5) (see also \cite{VH}, p.152).

If $X=D$ in (1.9), then the term $\int_S\sigma u\bar v dS$ is
absent in the definition of $[u,v]_\gamma$ in (3.2)
and assumptions (1.4) and (1.5) can be dropped.

If $X=N$ in (1.9), then the term $\int_S\sigma u\bar v dS$ is
also absent in the definition of $[u,v]_\gamma$ in (3.2),
 assumption (1.4) is used and assumption (1.5) is not used.

If $X=R$ in (1.9), then both  assumptions (1.4) and (1.5)
are used.

Assumption (1.4) allows one to conclude that (3.5) implies
(3.12) (see below) in the case of Neumann or Robin boundary
conditions.

Assumption (1.5) allows one to conclude that (3.5) implies
that the term  $\int_S\sigma \psi_n \bar \varphi \,dS$
converges and can be estimated by the $ H^1(D'_R)$-norm of
$\psi_n$ for any test function $\varphi$ in (3.14) (see below).

Assuming \num{1.1}--\num{1.3},
one has:
\begin{lemma}
The form $B_\gamma[\cdot,\cdot]$ is continuous
in $H^1(D')\times H^1(D')$ and for all sufficiently
large $\gamma>0$
there exist $\beta_j, j=1,2,$ such that
$$
\beta_1 \|u\|^2_{1}\leq B_\gamma[u,u]\leq \beta_2 \|u\|^2_{1} \qquad
\beta_1>0.
$$
\end{lemma}
{\bf Proof}: One has
$|\int_{B_R}q|u|^2dx|\leq c(R)||u||^2_{1}$ if \num{1.3} holds.
Indeed, if $u\in H^1(B_R)$, then, by the Sobolev embedding theorem,
one has $u\in L^{\frac {2n}{n-2}}(B_R)$. By H\"older's inequality,
$$
\left|\int_{B_R}q|u|^2dx\right|
\leq ||q||_{L^p(B_R)} ||u||^2_{L^{2p'}(B_R)}
,
$$
where $p':=\frac p {p-1}$. Choose $p'<\frac n{n-2}$. Then
$p>\frac n2$. If $p>\frac n2$ and $u\in H^1(B_R)$, then
$|\int_{B_R}q|u|^2dx|\leq c(R)||u||^2_{1}$, as claimed.
Moreover, since the embedding $H^1(D_{Rloc}) \to L^{2p'}(D_{Rloc})$ 
is compact
for $p'<\frac n{n-2}$, it follows from (1.3) and (1.4) that
$$\left|\int_{D_R}q|u|^2dx\right|\leq \nu ||u||^2_{H^1(D_R)}+C(\nu)
||u||^2_{L^{2}(D_R)}
$$
for any $\nu>0$, however small (see \cite{R4}). 
Assumption (1.5) implies
$$\left|\int_{S}\sigma|u|^2ds\right|
\leq \nu ||u||^2_{H^1(D_R)}+C(\nu)
||u||^2_{L^{2}(D_R)} $$
for any $\nu>0$, however small.
Lemma 3.1 is proved.  \bo

It follows from the above lemma that
the norm $[B_\gamma (u,u)]^{1/2}$ is equivalent to $H^1(D')$ norm,
the form $B_\gamma[\cdot,\cdot]$ is closed, symmetric
and densely defined in the Hilbert space $H:=L^2(D')$.
Therefore this form defines a unique self-adjoint 
nonnegative operator $L$
in $H=L^2(D')$ with domain dense in $H^1(D')$.
The points $z$ with $\Im z\neq 0$ do not belong to its spectrum.
Thus
problem \num{3.1}, with $\ep>0$ has  a unique solution in $H^1(D')$.
We have proved the following:
\begin{prop} \label{p3.1}
Problem \num{3.1} has a unique solution $w_\ep \in H^1(D')$.
\end{prop}
In the proof of the above Proposition we have not used
the compactness of the support of the potential $q(x)$ and gave
the proof valid for $q(x)$  satisfying \num{1.3}.

\subsection{The limiting absorption principle}

In the above subsection we have proved that for each $\ep>0$,
problem \num{3.1} admits a solution $w_\ep$.
In this subsection we shall prove that one can take the limit
for $\ep$ going to zero, and get the solution of \num{2.6}--\num{2.8}.

We first prove the following fundamental Lemma.
\begin{lemma} \label{l3.2}
Suppose
$\psi_n\in H^1_{loc}(D')\cap H^0_{-s}(D')$, with $s>1$,
and, in the weak sense,
\begin{eqnarray}
L\psi_n- \left( k^2+i\ep_n \right)\psi_n &=& h_n  \label{3.3} \\
\Gamma^{R}\psi_n  &=& 0 \; ,  \label{3.4}
\end{eqnarray}
with $\ep_n\downarrow 0$, $h_n\in L^2(D')$ have compact support
and $h_n\rightarrow h$ in $L^2(D')$, where $h$ has compact support.
Moreover suppose
\begin{equation}
\|\psi_n\|_{0,-s}\leq M \; , \label{3.5}
\end{equation}
where $M>0$  is a constant independent of $n$.
Then there exists a subsequence of  $\left\{\psi_n\right\}_{n\in\bN}$,
denoted again by  $\left\{\psi_n\right\}_{n\in\bN}$,
and a $\psi\in \Hloc{2}\cap H^1_{-s}(D')$, with $s>1$, such that:
\begin{eqnarray}
\psi_n\longrightarrow \psi \quad &\mbox{\rm{strongly in}}& \;
\Hloc{2} \; ,  \label{3.6} \\
\psi_n\longrightarrow \psi \quad &\mbox{\rm{strongly in}}& \;
H^0_{-s}(D') \; ,  \label{3.7}
\end{eqnarray}
with $\psi$ solving the limiting  problem:
\begin{eqnarray}
L\psi-   k^2 \psi &=& h,  \label{3.8} \\
\Gamma^{R}\psi  &=& 0 \; . \label{3.9}
\end{eqnarray}
Moreover
\begin{equation}
\lim_{r\rightarrow \infty}\int_{|x|=r} | \psi_r -ik\psi|^2ds =0. \;
\label{3.10}
\end{equation}
\end{lemma}
{\bf Proof:}

{\it Step 1: Convergence of $\psi_n$ in $\Lloc$ and weak convergence
in $H^1(D_r) \, \forall r>R$.}

Due to \num{3.5}, there exists a subsequence
$\psi_n \in H^0_{-s}(D')$ which
converges weakly in $L^2_{loc}(D')$:

\begin{equation}
\psi_n\longrightarrow \psi \quad \mbox{\rm{weakly in}} \quad \Lloc.
\label{3.11}
\end{equation}
Let us now prove that this subsequence $\psi_n$ is
bounded in $H^1_{loc}(D')$ uniformly with respect to $n$.
If this is proved, then one gets:
\begin{equation} \psi_n\longrightarrow \psi \quad \mbox{\rm{weakly
in}} \quad \Hloc{1},
\psi_n\longrightarrow \psi \quad \mbox{\rm{strongly
in}} \quad L^2(D_r) \forall r>R, \label{3.12}
\end{equation}
for a subsequence, denoted also by $\psi_n$.
To prove that $||\psi_n||_{H^1(D_r)}<c(r)\,\, \forall r>R$,
take in the weak formulation of (3.3), similar
to (3.1), the test function $\varphi:=\psi_n \beta(x)$,
where $\beta(x)$ is a cut-off function, $0\leq \beta(x) \leq 1$,
$\beta(x)=1$ for $|x|<r$, $\beta(x)=0$ for $|x|>r+1$, $r>R$.
It then follows from the analog of (3.1) (see (3.13) below) that
$||\psi_n||_{H^1(D_r)}<c(r)\,\, \forall r>R$,
where $c(r)>0$ is a constant independent of $n$, as claimed.

Similarly one can prove that, if $s>1$ then
$$||\psi_n-\psi||_{H^1_{-s}(D')}\to 0, \quad n\to \infty.$$
This is done as follows: write (3.1) with $w=\psi_n$ and subtract
from (3.1) with $w=\psi_m$. Use (1.1), (3.7), (3.12), (3.16)--(3.18)
to get the desired conclusion.

If $\psi_n$ is a weak solution of
the problem \num{3.3}--\num{3.4}, one can write:
\begin{equation}
\begin{array}{l}
\displaystyle
\int_{D'}
\left[a_{ij}\pardi \psi_n\pardj \bar{\vp} dx +
\left(q-k^2-i\ep_n\right)\psi_n\bar{\vp} \right] + \int_S \sigma
\psi_n\bar{\vp}ds =(h_n,\vp)\;
\\[1.2ex]
\qquad\qquad \forall \vp\in
\tilde{H}^1\left( D'\right)  \; .
\end{array}
\label{3.13}
\end{equation}
Using \num{3.11} and
\num{3.12} one can pass to the limit in the above equation and get:
\begin{equation}
\displaystyle
\begin{array}{l}
\displaystyle
\int_{D'} \left[a_{ij}\pardi \psi \pardj \bar{\vp} dx +
\left(q-k^2 \right)\psi \bar{\vp} \right] + \int_S \sigma \psi \bar{\vp}ds
=(h ,\vp)\;
\\[1.2ex]
\qquad\qquad \forall \vp\in \tilde{H}^1\left( D'\right) \; .
\end{array}
\label{3.14}
\end{equation}

\smallskip
{\it Step 2: Convergence of $\psi_n$ in $\Hloc{2}$.}

By the elliptic regularity, one concludes that
$\psi_n$ and $ \psi$ are in $\Hloc{2}$.
Moreover, due to the density of $\tilde{H}^1(D')$ in $L^2(D')$,
it follows from  \num{3.13} and \num{3.14} that $\psi_n$ and
$\psi$ satisfy (almost everywhere in $D'$) \num{3.3} and
\num{3.8} respectively.

Subtract \num{3.8} from \num{3.3}, and use the elliptic
estimate to get:
\begin{equation}
\begin{array}{l}
\displaystyle
\|\psi_n-\psi\|_{H^2(D_1)}\leq c \left( \|\psi_n-\psi\|_{L^2(D_2)}+
\|h_n-h\|_{L^2(D_2)} +\ep_n\|\psi_n\|_{L^2(D_2)}\right)
\\[1.2ex]
\qquad\qquad \forall D_1\subset \subset D_2\subset \subset D' \; ,
\end{array}
 \label{3.15}
\end{equation}
where $c>0$ is a constant which depends on $D_1$ and $D_2$
but not on $n$.
This estimate  proves  \num{3.6},
since $\ep_n\to 0$ and $||\psi_n||_{L^2(D_2)}$
is bounded.

\smallskip
{\it Step 3: The representation formula and the radiation condition.}

If $\psi_n$ satisfies \num{3.3}, assumption (1.2) holds
and $q(x)=0$ for $|x|>R$, then the following
representation  formula holds:
\begin{equation}
\psi_n(x)= \int_{S_R} \left[ \psi_n(s) \pardN g_{\ep_n}(x,s) -
g_{\ep_n}(x,s)  \pardN \psi_n(s)  \right]ds
\qquad \mbox{for} \quad x\in B'_R\;
\; ,  \label{3.16}
\end{equation}
where
$\displaystyle
g_{\ep_n}=\frac{e^{i\sqrt{k^2+i\ep_n}|x-y|}}{4\pi|x-y|}$
 is the Green function of the operator
$\Delta+ (k^2+i\ep_n)$
and $N$ in \num{3.16} and below stands for the normal
to $S_R$ pointing into $B'_R$.
One can pass to the limit in \num{3.16} (due to the convergence
of $\psi_n$ in $\Hloc{2}$ proved in Step 2),
and get the following representation
formula for $\psi$:
\begin{equation}
\psi(x)= \int_{S_R} \left[ \psi(s) \pardN g(x,s) -
 g (x,s) \pardN \psi(s) \right] ds
\; ,   \label{3.17}
\end{equation}
where
$\displaystyle
g=\frac{e^{ik|x-y|}}{4\pi|x-y|}$
is the Green function of the operator
$\Delta+k^2$ in $\bR^3$.
Equation \num{3.17} implies that $\psi$ satisfies the radiation
condition \num{3.10}.

\smallskip
{\it Step 4: A uniform estimate of the behavior of $\psi_n$ at infinity.}

The representation formulas \num{3.16} for $\psi_n$ and \num{3.17}
for $\psi$, imply the following uniform estimate for the behavior
of $\psi_n$ at infinity (in  $R^3$):
\begin{equation}
\sup_n {\left(|\psi_n|+|\nabla\psi_n|\right)}\leq \frac{c}{|x|}
\qquad \mbox{for}\quad x\in B'_R \; , \label{3.18}
\end{equation}
where $c>0$ is a constant independent of $x\in B'_R $.
This estimate will be crucial in the next Step.
Note that (3.12), (3.18) and the estimate
$||\psi_n||_{H^1(D_r)}\leq c(r)\,\, \forall r\geq R$
imply that $\psi \in H_{-s}^1(D'),\,\, s>1$.

\smallskip
{\it Step 5: Convergence  of $\psi_n$ in $H^0_{-s}(D')$ and
the conclusion of the proof.}

To complete the proof of the Lemma we have to prove the convergence
property \num{3.7}.
Estimate \num{3.18} and the assumption $s>1$ imply:
\begin{equation}
\begin{array}{l}
\displaystyle
\|\psi_n-\psi\|^2_{0,-s}=
\int_{D'}\frac{|\psi_n-\psi|^2}{(1+|x|^2)^{s/2}}dx
\\[2.3ex]
\displaystyle
\qquad\qquad\qquad =
\int_{B_R\cap D'}\frac{|\psi_n-\psi|^2}{(1+|x|^2)^{s/2}}dx +
\int_{D'_R}\frac{|\psi_n-\psi|^2}{(1+|x|^2)^{s/2}}dx
\\[2.3ex]
\displaystyle
\leq \sup_{B_R\cap D'}{\left[\frac{1}{(1+|x|^2)^{s/2}} \right]}
\|\psi_n-\psi\|^2_{L^2(B_R\cap D')}+4\pi c^2   \int_R^\infty
\frac{1}{r^2}\frac{1}{(1+r^2)^{s/2}}  r^2 dr
\\[1.2ex]
\displaystyle
\leq \eta \; ,
\end{array}
\end{equation}
with $\eta>0$ arbitrarily small.
In the last step of the above estimate we have chosen $R$  so that
$4\pi c^2\int_R^\infty  dr/(1+r^2)^{s/2} \leq \eta/2$, and \num{3.11}
has been used.
This proves \num{3.7} in $\bR^3$. In  other space dimensions
the proof is analogous.

\vskip.1cm
With the help of   Lemma~\ref{l3.2}  we prove the following {\it a priori}
estimate for the  solutions $w_\ep$ of the problem \num{3.1}.
In estimate \num{3.20}    below we take $0<\ep<1$, but we could take
$0<\ep<\ep_0$,
where $\ep_0>0$ is an arbitrary small fixed number.

\begin{prop} \label{p3.2}
The solution $w_\ep$ of problem \num{3.1} satisfies the following
a priori estimate:
\begin{equation}
\sup_{0<\ep<1}\|w_\ep\|_{0,-s}\leq c \; . \label{3.20}
\end{equation}
\end{prop}
{\bf Proof:}

We prove this Proposition by contradiction. Suppose \num{3.20}
is false. Find $\ep_n\downarrow 0$ such that
$$
\|w_{\ep_n}\|_{0,-s} \geq n \; .
$$
Define $\psi_n:=w_{\ep_n}/\|w_{\ep_n}\|_{0,-s}$.
Clearly $\psi_n$ satisfy all the hypotheses of Lemma~ \ref{l3.2}
with $h_n=f/\|w_{\ep_n}\|_{0,-s}$, $h_n \rightarrow 0$ in $L^2(D')$.
Lemma \ref{l3.2} yields that $\psi_n\rightarrow \psi$ where
$\psi$ solves the  problem:
\begin{eqnarray}
&&L\psi-   k^2 \psi = 0,  \nonumber \\[0.5ex]
&&\Gamma^{R}\psi  = 0, \; . \nonumber \\[0.5ex]
&&\lim_{r\rightarrow \infty}\int_{|x|=r} | \psi_r -ik\psi|^2ds =0
\, .  \nonumber
\end{eqnarray}
By the uniqueness Theorem \ref{th2.1} we get $\psi\equiv0$.
But this contradicts to the fact that $\|\psi_n-\psi\|_{0,-s}
\rightarrow 0$ and $\|\psi_n\|_{0,-s}=1$. Proposition~\ref{p3.2} is
proved.
\bo
\vskip 5mm

If  $w_\ep$ satisfies \num{3.20} one can apply the
Lemma~ \ref{l3.2}
 to $\psi_n=w_{\ep_n}$ for some $\ep_n\downarrow 0$,
and with $h_n=f$.
We have therefore proved the existence of a $w$ solving
the problem \num{2.6}--\num{2.8} and the main result of this Section:

\begin{theorem} \label{th3.1}
Scattering problem \num{1.sc}--\num{1.ra} with a compactly supported
potential $q$ has  a unique solution $u$ of the form $u=w+\zeta u_0$.
Here $w\in \Hloc{2}\cap H^1_{-s}(D')$, with $s>1$, and $\zeta$ is a
function defined above formula \num{2.6}.
\end{theorem}

\section{Existence for a decaying potential}
\setcounter{equation}{0}

In this section we  prove that the scattering problem
\num{1.sc}--\num{1.ra}, with a potential $q\in H^1_{s}$ has
a solution.
We shall prove  the existence of the solution in the same way
we did for the scattering problem with a compactly supported
potential.
First we look for a solution $u$ of the form $u=w+\zeta u_0$
so that the problem is reduced to
\num{2.6}--\num{2.8}.
Then we prove the existence of a unique solution $w_\ep$ for
scattering problem \num{3.1}.
Finally we shall prove that one can take the limit
$\ep\downarrow 0$ and get the solution of \num{2.6}--\num{2.8}.

\subsection{Existence for the equation with absorption}

\begin{prop} \label{p4.1}
Problem \num{3.1}, with $\ep>0$, has a unique solution
$w_\ep \in H^1(D')$.
\end{prop}
The proof of this proposition is the same
as the one of Proposition \ref{p3.1}.

\subsection{The limiting absorption principle}

In this subsection we prove that the solution $w_\ep$ of
problem \num{3.1} converges, as $\ep\downarrow 0$,    to the  solution of
\num{2.6}--\num{2.8}.

The main step is the proof of  Lemma 4.3
 below.
We first state two Lemmas:

\begin{lemma} \label{l4.1}
Suppose that $|g(x,y)|\leq c|x-y|^{-1}$, and $|f(x)|\leq
c\left( 1+|x|^2\right)^{-s/2}$,  $x,y\in D'$ and  $s>3$.
Then
\begin{equation}
\leftab \int_{B'_R} g(x,y) f(y) dy\rghtab +
\leftab \int_{B'_R} \nabla_{x} g(x,y) f(y) dy\rghtab\leq \frac{c}{1+|x|}
\; ,
\quad x\in D'.
\end{equation}
\end{lemma}
{\bf Proof:}

Let  $r=|y|$, $\rho=|x|$, $y=r\left( \sin{\theta}\cos{\phi},
 \sin{\theta}\sin{\phi},
\cos{\theta}\right)$,
 $u=\cos{\theta}$, and let  $r,\theta$ and $\phi$ be
the  spherical coordinates. One has
\begin{eqnarray}
\leftab \int_{B'_R} g(x,y) f(y) dy\rghtab &\leq&
\int_{B'_R\cap \left\{y\,:\;|y|\leq\rho/2\right\}}
\frac{|f(y)|}{|x-y|} dy
+  \int_{ \left\{y\,:\;|y|\geq\rho/2\right\}}\frac{|f(y)|}{|x-y|} dy
\nonumber \\[0.5ex]
&\leq& \frac{c }{|x|}\int_{B'_R}|f(y)|dy +
c\int_{ \left\{y\,:\;|y|\geq\rho/2\right\}}
\frac{dy}{|x-y| \left( 1+|y|^2    \right)^{s/2}}
\nonumber \\[0.5ex]
&\leq &\frac{c }{|x|}+
2\pi c\int_{\rho/2}^\infty dr
\frac{r^2}{\left( 1+r^2    \right)^{s/2}}
\int_{-1}^{1}\frac{du}
{\left(\rho^2+r^2 -2r\rho u   \right)^{1/2}} 
\nonumber \\[0.5ex]
&\leq & \frac{c }{|x|}+
c_1\int_{\rho/2}^\infty\frac{dr}{\left(1+r^2\right)^{\frac{s-2}{2}}}
\frac{1}{\min(r,\rho)}
\,\leq \, \frac{c }{|x|}. \nonumber
\end{eqnarray}
The second integral in (4.1) can be estimated similarly.

If $s>2$ in lemma 4.1, then the argument above
yields $o(1)$ as $|x|\to \infty$, in place of $\frac{c }{|x|}$ term.
By a similar argument one can prove that if $s>2$ and $g=\frac
{e^{ik|x-y|}}{4\pi |x-y|}$, then the function
$h:=\int_{B'_R} g(x,y) f(y)dy $ satisfies the radiation condition:
$|x||\frac {\partial h}{\partial |x|}-ikh|\to 0$ as $|x|\to \infty$
uniformly in the directions of $x$. This remark does not allow
one to replace $s>3$ in the assumption (1.3) by
$s>2$. The reason is: if $q$ is not compactly supported, the function
$f$ defined by (2.6) contains the term $q(x)\zeta(x)u_0$ which
decays as $O((1+|x|^2)^{-\frac s2})$ for large $|x|$.
If $2<s<3$, then the argument given in lemma 4.1 is
not sufficient for getting estimate (4.1). It is
probable that the basic result, Theorem 4.1 below,
can be established for $s>2$ in (1.3), but some additional
argument is needed for a proof of such a result.

\begin{lemma} \label{l4.2}
Suppose that $|g(x,y)|\leq c|x-y|^{-1}$ and $|f(x)|\leq
c\left( 1+|x|^2\right)^{-s/2}$,
$x,y\in D'$,  and $\psi\in H^0_{-s}(D')$    with $s>3$.
Then
\begin{equation}
\leftab \int_{B'_R} g(x,y) f(y) \psi(y) dy \rghtab
+\leftab \int_{B'_R} \nabla_{x} g(x,y) f(y) \psi(y) dy \rghtab\leq
\frac{c}{|x|}
\;,\quad x\in D'
.
\end{equation}
\end{lemma}
{\bf Proof:}

Denote $|x|:=\rho,\, |y|:=r$,
$ \;
 T(x,y):=
 |x-y|^{-2}(1+|y|^2)^{-s/2}\,
$.
One has:
\begin{eqnarray}
&&\leftab \int_{B'_R} g(x,y) f(y) \psi(y) dy\rghtab 
\nonumber \\[0.5ex]
&&\leq \left[\int_{B'_R} |g(x,y) f(y)|^2
\left(1+|y|^2\right)^{s/2}dy\right]^{1/2}
\left[\int_{B'_R} |\psi(y)|^2
\left(1+|y|^2\right)^{-s/2}dy\right]^{1/2} 
\nonumber \\[0.5ex]
&&\leq  c  \|\psi\|_{0,-s}\left[\int_{B'_R}
T(x,y)\,dy
\right]^{1/2} 
\nonumber \\[0.5ex]
&&\leq
 c_1\left\{ \left[  \int_{B'_R\cap \left\{y\,:\;|y|\leq\rho/2\right\}}
T(x,y)\,dy
\right]^{1/2}+
\left[\int_{ \left\{y\,:\;|y|\geq\rho/2\right\}}
T(x,y)\,dy
\right]^{1/2}
\right\} 
\nonumber \\[0.5ex]
&&\leq
\frac{c}{|x|} +2\pi c \left[ \int_{\rho/2}^\infty dr
\frac{r^2}{\left( 1+r^2
   \right)^{s/2}}
\int_{-1}^{1}\frac{du}{\rho^2+r^2 -2r\rho u   }
\right]^{1/2} 
\;\leq\;
\frac{c}{|x|} \, .\nonumber
\end{eqnarray}
The second integral in (4.2) can be estimated similarly.

We can now prove  a Lemma analogous to  Lemma \ref{l3.2} of the
previous Section.

\begin{lemma} \label{l4.3}
Suppose
$\psi_n\in H^1(D')\cap H^0_{-s}(D')$,
with $s>3$, and, in the weak sense,
\begin{eqnarray}
L\psi_n- \left( k^2+i\ep_n \right)\psi_n &=& h_n,  \label{4.3} \\
\Gamma^{R}\psi_n  &=& 0 \; ,  \label{4.4}
\end{eqnarray}
where  $\ep_n\downarrow 0$, $h_n\in L^2(D'),$
$|h_n(x)|\leq c(1+|x|^2)^{-s/2}$ , $s>3$,
and $h_n\rightarrow h$ in $L^2(D')$, where $|h(x)|\leq c(1+|x|^2)^{-s/2}$.
Moreover, suppose
\begin{equation}
\|\psi_n\|_{0,-s}\leq M \; ,\quad  s>3\, , \label{4.5}
\end{equation}
where $M$ is a constant independent of $n$.
Then there exists a subsequence of  $\left\{\psi_n\right\}_{n\in\bN}$,
denoted again by $\left\{\psi_n\right\}_{n\in\bN}$,
and a $\psi\in \Hloc{2}\cap H^1_{-s}(D')$, such that:
\begin{eqnarray}
\psi_n\longrightarrow \psi \quad &\mbox{\rm{in}}& \quad
\Hloc{2} \; ,  \label{4.6} \\
\psi_n\longrightarrow \psi \quad &\mbox{\rm{in}}& \quad
H^0_{-s}(D') \; ,  \label{4.7}
\end{eqnarray}
where  $\psi$ solves the following problem:
\begin{eqnarray}
L\psi-   k^2 \psi &=& h\, ,  \label{4.8} \\
\Gamma^{R}\psi  &=& 0 \; ,\label{4.9}
\end{eqnarray}
\begin{equation}
\lim_{r\rightarrow \infty}\int_{|x|=r} | \psi_r -ik\psi|^2ds =0 \;.
\label{4.10}
\end{equation}
\end{lemma}
{\bf Proof:}

{\it Step 1: Convergence of $\psi_n$ in $\Lloc$ and weak convergence
of $\psi_n$ in
$H^1(D_r)$, $r\geq R$.}

If (\ref{4.5}) holds, then
\begin{eqnarray}
\psi_n\longrightarrow \psi &\quad \mbox{\rm{strongly in}} &\quad
\Lloc \,,\label{4.11}  \\
\psi_n\longrightarrow \psi &\quad \mbox{\rm{weakly in}} &\quad
 H^1(D_r)\, \quad \forall r\geq R \label{4.12}  \; .
\end{eqnarray}
This is  proved in Lemma \ref{l3.2}.

\smallskip
{\it Step 2: Convergence  $\psi_n$ in $\Hloc{2}$.}

One has
\begin{equation}
\psi_n\longrightarrow \psi \quad \mbox{\rm{strongly in}} \quad
\Hloc{2} \label{4.13} \; .
\end{equation}
This is  proved as in Lemma \ref{l3.2}.

As in the Section 3,
$\psi_n$ satisfies \num{4.3} and \num{4.4}, and $\psi$ satisfies
\num{4.8} and \num{4.9}.

\smallskip
{\it Step 3: The representation formula for
$\psi_n$ and the radiation condition.}

If $\psi_n$ satisfies \num{4.3} then  the following
representation  formula holds for $x\in B'_R$:
\begin{eqnarray}
\psi_n(x)&=&
\int_{S_R} \left[ \psi_n(s) \pardN g_{\ep_n}(x,s) -
\pardN \psi_n(s) g_{\ep_n}(x,s) \right]ds \nonumber \\
&&+  \int_{B'_R} g_{\ep_n}(x,y) h_n(y) dy
-  \int_{B'_R} g_{\ep_n}(x,y) q(y) \psi_n(y)dy
\; ,  \label{4.14}
\end{eqnarray}
where $N$ in the above formula and below denotes the normal
to $S_R$ pointing into $B'_R$.
Since  $\psi_n$ converges in $\Hloc{2}$ one can pass to the limit
in (\ref{4.14})  and get:
\begin{eqnarray}
\psi(x)&=&
\int_{S_R} \left[ \psi(s) \pardN g(x,s) -
\pardN \psi(s) g(x,s) \right]ds \nonumber \\
&&+  \int_{B'_R} g(x,y) h(y) dy
-  \int_{B'_R} g(x,y) q(y) \psi(y)dy
\; .  \label{4.15}
\end{eqnarray}
Equation \num{4.15} implies that $\psi$ satisfies
the radiation condition.

\smallskip
{\it Step 4: A uniform estimate of the behavior of  $\psi_n$ at infinity.}

As in the case of a compactly supported potential one gets the
following estimate for the behavior of $\psi_n$ at
infinity:
\begin{equation}
\sup_n {\left(|\psi_n|+|\nabla\psi_n|\right)}\leq \frac{c}{|x|}
\qquad \mbox{for}\quad x\in B'_R \; , \label{4.16}
\end{equation}
where $c>0$ is a constant independent of $x\in B'_R$.
This follows from \num{4.14}
and \num{4.15}.
The additional terms which  appear because the potential and
the source term are not compactly supported, are  estimated in
Lemmas \ref{l4.1} and \ref{l4.2}. Estimates (4.12) and (4.16)
imply $\psi\in H^1_{-s}(D'),\,\, s>1$.

\smallskip
{\it Step 5: Convergence of  $\psi_n$ in $H^0_{-s}(D')$ and
the conclusion of the proof.}

One  proves that $\|\psi_n-\psi\|_{0,-s}\rightarrow 0$ as in Lemma
\ref{l3.2}.  This concludes the proof
of  Lemma  \ref{l4.3}.
\bo
\vskip.2cm

Now one gets the following {\it a priori} estimate for
the solution  $w_\ep$ of the problem \ref{3.1}:

\begin{prop}\label{p4.2}
The solution of  problem \num{3.1} satisfies the following a priori
estimate:
\begin{equation}
\sup_{0<\ep<1}\|w_\ep\|_{0,-s}\leq c \; ,\quad s>1\,.  \label{4.17}
\end{equation}
\end{prop}
The proof is based on Lemma \ref{l4.3}, and is the same as
the proof of Proposition \ref{p3.2}.

Let us state {\it the main result of this paper}, whose proof is
based on the {\it a priori} estimate \num{4.17} and  Lemma
\ref{l4.3}:

\begin{theorem} \label{th4.1}
Assume that conditions (1.1)--(1.5) hold and $\sigma(s)$
is a real-valued $L^{\infty}(S)$- function. Then
the scattering problem \num{1.sc}--\num{1.ra} has  a
weak solution $u$ of the form $u=w+\zeta u_0$,
with $w\in \Hloc{2}\cap H^1_{-s}(D')$\,,  $\,\,s>1$,
$w$ satisfies (3.1) and (2.8), $\zeta$ is defined
above formula (\ref{2.6}), and this solution is unique in the above
space.
\end{theorem}


\end{document}